\documentclass[%
 reprint,
amsmath,amssymb,
aps,
prl,
]{revtex4-1}

\usepackage{graphicx}
\usepackage{dcolumn}
\usepackage{bm}
\begin{document}

\preprint{AIP/123-QED}

\title{Lensless Wiener-Khinchin telescope based on high-order spatial autocorrelation of thermal light}

\author{Zhentao Liu, Xia Shen, Honglin Liu, Hong Yu, Shensheng Han*}

\affiliation{Shanghai Institute of Optics $\&$ Fine Mechanics, Chinese Academy of Sciences, Shanghai 201800, China\\
*Corresponding author: sshan@mail.shcnc.ac.cn}



\date{\today}

\begin{abstract}
The resolution of a conventional imaging system based on first-order field correlation can be directly obtained from the optical transfer function. However, it is challenging to determine the resolution of an imaging system through random media, including imaging through scattering media and imaging through randomly inhomogeneous media, since the point-to-point correspondence between the object and the image plane in these systems cannot be established by the first-order field correlation anymore. In this paper, from the perspective of ghost imaging, we demonstrate for the first time to our knowledge that the point-to-point correspondence in these imaging systems can be quantitatively recovered from the high-order correlation of light fields, and the imaging capability, such as resolution, of such imaging schemes can thus be derived by analyzing high-order correlation of the optical transfer function. Based on this theoretical analysis, we propose a lensless Wiener-Khinchin telescope based on high-order spatial autocorrelation of thermal light, which can acquire the image of an object by a snapshot via using a spatial random phase modulator. As an incoherent imaging approach illuminated by thermal light, lensless Wiener-Khinchin telescope can be applied in many fields such as X-ray astronomical observations.
\end{abstract}

\maketitle

Imaging resolution is an important metric of various imaging systems, including microscopy, astronomy, and photography \cite{sheppard2017resolution, baker2002limits, Roggemann1997Improving, Hao2013microscopy}. It is well known that the operating wavelength $\lambda$ and the aperture $D$ of an imaging system are two key parameters for resolution \cite{sheppard2017resolution, baker2002limits}. Generally speaking, in conventional imaging systems, where a point-to-point correspondence between the object and the image plane can be established based on first-order field correlation, the resolution can be directly analyzed from a transmission function of the imaging system, and it is proportional to $\lambda/D$. Therefore, a shorter wavelength $\lambda$ and/or a larger aperture $D$ is required for a higher resolution, however, a large aperture leads to demanding requirements on the manufacture of a traditional monolithic optical telescope, which is arduous especially for X-ray imaging.

Recently, emerging systems through random media \cite{Choi2011Overcoming, Liutkus2014Imaging, Newman2014Imaging, Yilmaz2015Speckleb, liu2016spectral, Sahoo2017Single, Antipa2017DiffuserCam, Wang2015Ultra, Bertolotti2012Non, Katz2012Looking, katz2014non, Yang2014Imaging, labeyrie1970attainment, Gezari1972Speckle} have been built, which include imaging through scattering media and imaging through randomly inhomogeneous media. However, it is challenging to determine the resolution of these imaging systems, since the point-to-point correspondence between the object and the image plane in these systems cannot be established by the first-order correlation anymore. In this paper, we show that when the statistical properties of the random media are known as a prior, the resolution of such imaging system can be deduced by analyzing high-order correlation of light fields from the prospective of ghost imaging (GI) \cite{Liu2007Ghost, Zhang2009Improving, liu2016spectral}. Based on this theoretical analysis, a lensless Wiener-Khinchin telescope is further proposed based on high-order spatial autocorrelation of thermal light, which can acquire the image of the object in a single shot by using a spatial random phase modulator (SRPM). We theoretically and experimentally demonstrate that different from conventional imaging systems, the resolution of lensless Wiener-Khinchin telescope not only depends on the aperture of SRPM, but also on the statistical properties of it. The influence of signal bandwidth is also investigated. Moreover, experimental results for both far away and equivalent infinity far away imaging proves the feasibility of the proposed lensless Wiener-Khinchin telescope in astronomical observations.

\begin{figure}[htbp]
\centering
\includegraphics[width=0.8\linewidth]{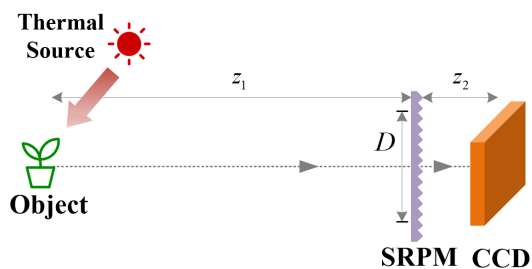}
\caption{Schematic of a lensless Wiener-Khinchin telescope. $D$ is the diameter of the SRPM. $z_1$ and $z_2$ are distances from the object and detection planes to the SRPM, respectively.}\label{fig:Schematic}
\end{figure}

The proposed lensless Wiener-Khinchin telescope (Fig. \ref{fig:Schematic}) consists of a SRPM and a charge-coupled device (CCD) detector, which detects the intensity distribution of the modulated light field. The object is illuminated by a thermal light source.

For incoherent imaging \cite{goodman2005introduction}, the spatial intensity distribution detected by the CCD detector is

\begin{equation}
I_t(r)=\int_{-\infty}^{\infty}{I_0(r_0)h_I(r;r_0)dr_0}\quad ,
\label{eq3}
\end{equation}
where $I_0(r_0)$ is the intensity distribution in the object plane, $h_I(r;r_0)=h_E(r;r_0)h_E^*(r;r_0^\prime)$ is the incoherent intensity impulse response function, $h_E(r;r_0)$ is the point-spread function (PSF) of the imaging system.

Considering the spatial random phase modulator for thermal light as an ergodic process, the second-order spatial autocorrelation of the measured light field is \cite{cheng2004incoherent, Shapiro2012physics}
\begin{equation}
\begin{split}
&G_{{I_t}}^{(2)}\left( {r + \Delta r,r} \right)\\
&=\left\langle {E_{t}^*\left( {r + \Delta r} \right)E_{t}^*\left( r \right){E_t}\left( r \right){E_t}\left( {r + \Delta r} \right)} \right\rangle_{r}\\
&=\overline{\left\{E_{t}^*\left( {r + \Delta r} \right)E_{t}^*\left( r \right){E_t}\left( r \right){E_t}\left( {r + \Delta r} \right)\right\}}_{s}\\
&=\overline{\left\{{I_t}\left( r \right){I_t}\left( {r + \Delta r} \right)\right\}}_{s}
\end{split}
\label{eq4}
\end{equation}
where $\left\langle\cdot\right\rangle_{r}$ is the spatial average over the coordinate $r$, and $\overline{\{\cdot\}}_{s}$ is the ensemble average of the SRPM. Plugging Eq. (\ref{eq3}) into Eq. (\ref{eq4}), we have
\begin{equation}
\begin{split}
G_{{I_t}}^{(2)}\left( {r + \Delta r,r} \right)=\iint_{-\infty}^{\infty}&G_{h}^{(2)}(r+\Delta r, r_0 + \Delta r_0;r,r_0)\\
&I_0(r_0+\Delta r_0)I_0(r_0)dr_0 d\Delta r_0,
\end{split}
\label{eq5}
\end{equation}
where
\begin{small}
\begin{equation}
\begin{split}
&G_{h}^{(2)}(r+\Delta r, r_0 + \Delta r_0;r,r_0)=\\
&\overline {\left\{h_E^*(r+\Delta r;r_0+ \Delta r_0)h_E^*(r;r_0)h_E(r;r_0)h_E(r+\Delta r;r_0+ \Delta r_0)\right\}}_{s}\\
\end{split}
\label{eq6}
\end{equation}
\end{small}
is the second-order correlation function of PSFs.

According to the Central Limit Theorem \cite{goodman2007speckle9}, the light field $h_E(r;r_0)$ through the spatial random phase modulator obeys the complex circular Gaussian distribution in spatial domain \cite{goodman2007speckle9}, and $G_{h}^{(2)}(r+\Delta r,r_0+ \Delta r_0;r,r_0)$ can be written as \cite{goodman2015statistical44}
\begin{equation}
\begin{split}
&G_{h}^{(2)}(r+\Delta r,r_0+ \Delta r_0;r,r_0)\\
&=B \left[1+g_{h}^{(2)}(r+\Delta r,r_0+ \Delta r_0;r,r_0)\right]
\end{split}
\label{eq7}
\end{equation}
with $B=\overline{\left\{h(r;r_0)\right\}}_{s} \overline{\left\{h(r+\Delta r;r_0+ \Delta r_0)\right\}}_{s}$, and
\begin{equation}
\begin{split}
&g_{h}^{(2)}(r+\Delta r,r_0+ \Delta r_0;r,r_0)\\
&=\dfrac{\left|\overline{\left\{h_E^*(r+\Delta r;r_0+ \Delta r_0)h_E(r;r_0)\right\}}_{s}\right|^2}{B}\\
\end{split}
\label{eq8}
\end{equation}
is the normalized second-order correlation of PSFs.

For Fresnel diffraction, the PSF of a lensless Wiener-Khinchin telescope is
\begin{small}
\begin{equation}
\begin{split}
&h_E(r;r_0) = \dfrac{\exp\{{j2\pi (z_1+z_2)/\lambda}\}}{-\lambda^2 z_1 z_2}\exp\left\{{\dfrac{j\pi(r-r_0)^2}{\lambda (z_1+z_2)}}\right\}\\
&\int_{-\infty}^{\infty} P(r_m)t(r_m)\exp\left\{\frac{j\pi(z_1+z_2)}{\lambda z_1z_2}\left[r_m-\frac{z_1r+z_2r_0}{z_1+z_2}\right]^2\right\}dr_m,
\end{split}
\label{eq9}
\end{equation}
\end{small}
where $P(r_m)$ and $t(r_m)=\exp{\left[j2\pi(n-1){\eta(r_m)}/{\lambda}\right]}$ are the pupil function and the transmission function of SRPM, respectively, while $\eta(r_m)$ and $n$ are its height and refractive index, respectively. The space translation invariance of system in space (also known as memory effect \cite{Feng1988Correlations, Osnabrugge2017Generalized}) is required for the Fresnel approximation \cite{goodman2005introduction66}. Since the target of a telescope is very small compared with the imaging distance, the memory effect is satisfied in lensless Wiener-Khinchin telescope.

In general, the height ensemble average ${R_\eta }\left( {{r_m},{r_n}} \right)$ of SRPM obeys the following mathematical form \cite{Sinha1988X}
\begin{equation}
\begin{split}
&{R_\eta }\left( {{r_m},{r_n} } \right)=\overline{\left\{\eta \left( {{r_m}} \right)\eta \left( {{r_n} } \right)\right\}}_{s} \\
&= {\omega ^{\rm{2}}}\exp \left\{ { - {{\left( {\frac{{{r_m} - {r_n} }}{\zeta }} \right)}^2}} \right\}={R_\eta }\left( {\Delta {r_m}} \right),\quad \Delta {r_m}{\rm{ = }}{r_m} - {r_n},
\end{split}
\label{eq11}
\end{equation}
where $\omega$ and $\zeta$ are the height standard deviation and the transverse correlation length of SRPM, respectively. Substituting Eq. (\ref{eq9}) and Eq. (\ref{eq11}) into Eq. (\ref{eq8}) yields (see \href{link}{Supplement 1} for details)
\begin{equation}
\begin{split}
&g_{h}^{(2)}(r+\Delta r,r_0+\Delta r_0;r,r_0)\\
&\approx \left|\left\{{\exp \left\{-2\left[\dfrac{2\pi (n-1)}{\lambda}\right]^2\left[\omega^2-R_\eta\left(\dfrac{2\lambda z_1 z_2}{z_1+z_2}\nu\right)\right]\right\}}\right.\right.\\
&\left.\left.\otimes\mathcal{F}\left\{{\left| P\left(\mu+\mu_0\right)\right|^2}\right\}_{\mu \to \nu}\right\}_{\dfrac{z_1\Delta r+z_2\Delta r_0}{2\lambda z_1z_2}}\right|^2\\
&=g_{h}^{(2)}\left(\dfrac{z_1\Delta r+z_2\Delta r_0}{2\lambda z_1z_2}\right),
\end{split}
\label{eq12}
\end{equation}
where $\mathcal{F}\{\cdots\}_{\mu\to\nu}$ represents the Fourier transform of the function with the variable $\mu$ and the transformed function variable is $\nu$, and
\begin{equation}
\mu_0=\dfrac{z_1(2r+\Delta r)+z_2\left(2r_0+\Delta r_0\right)}{2\left(z_1 +z_2\right)}.
\label{eq13}
\end{equation}
 
Taking Eqs. (\ref{eq7}) and (\ref{eq12}) into Eq. (\ref{eq5}), we have
\begin{small}
\begin{equation}
\begin{split}
G_{{I_t}}^{(2)}\left( {r + \Delta r,r} \right)\approx B\left\{\left[1+g_{h}^{(2)}\left(\dfrac{\Delta r_0}{2\lambda z_1}\right)\right]\otimes{G_{{I_0}}^{(2)}\left( {r_0 + \Delta r_0,r_0} \right)}\right\}_{-\dfrac{z_1}{z_2}\Delta r},
\end{split}
\label{eq14}
\end{equation}
\end{small}
where
\begin{equation}
\begin{split}
G_{{I_0}}^{(2)}\left( {r_0 + \Delta r_0,r_0} \right)
=\int_{-\infty}^{\infty} {I_0(r_0)I_0\left(r_0+\Delta r_0\right)dr_0},
\end{split}
\label{eq15}
\end{equation}
and
\begin{small}
\begin{equation}
\begin{split}
&g_{h}^{(2)}\left(\dfrac{\Delta r_0}{2\lambda z_1}\right)\\
&=\left|\left\{{\exp \left\{-2\left[\dfrac{2\pi (n-1)}{\lambda}\right]^2\left[\omega^2-R_\eta\left(\dfrac{2\lambda z_1 z_2}{z_1+z_2}\nu\right)\right]\right\}}\right.\right.\\
&\left.\left.\otimes\mathcal{F}\left\{{\left| P\left(\mu+\mu_0\right)\right|^2}\right\}_{\mu \to \nu}\right\}_{\dfrac{\Delta r_0}{2\lambda z_1}}\right|^2.
\end{split}
\label{eq16}
\end{equation}
\end{small}

According to the Wiener-Khinchin theorem for deterministic signals \cite{Cohen1998generalization}(also known as autocorrelation theorem \cite{goodman2005introduction8}), we have
\begin{small}
\begin{equation}
\begin{split}
&G_{{I_0}}^{(2)}\left( {r_0 + \Delta r_0,r_0} \right)=\mathcal{F}^{-1}\left\{{\left|{\mathcal{F}\left\{{I_0(r_0 )}\right\}_{r_0\to f_{0}}}\right|^2}\right\}_{f_{0}\to \Delta r_0}.
\end{split}
\label{eq17}
\end{equation}
\end{small}

Substituting Eq. (\ref{eq17}) into Eq. (\ref{eq14}), we obtain
\begin{small}
\begin{equation}
\begin{split}
G_{{I_t}}^{(2)}&\left( {r + \Delta r,r} \right)
\propto \left\{\left[1+g_{h}^{(2)}\left(\dfrac{\Delta r_0}{2\lambda z_1}\right)\right]\right.\\
&\left.\otimes{\mathcal{F}^{-1}\left\{{\left|{\mathcal{F}\left\{{I_0(r_0)}\right\}_{r_0\to f_{0}}}\right|^2}\right\}_{f_{0}\to \Delta r_0}}\right\}_{-\dfrac{z_1}{z_2}\Delta r}.
\end{split}
\label{eq18}
\end{equation}
\end{small}

Eq. (\ref{eq18}) indicates that the energy spectral density ${\left|{\mathcal{F}\left\{{I_0(r_0)}\right\}_{r_0\to f_{0}}}\right|^2}$ of the intensity distribution $I_0(r_0)$ on the object plane can be separated from $G_{{I_t}}^{(2)}\left( {r + \Delta r,r} \right)$, and the resolution is determined by $g_{h}^{(2)}\left(\dfrac{\Delta r_0}{2\lambda z_1}\right)$. The image of ${I_0}(r_0)$ can be reconstructed by utilizing phase retrieval algorithms \cite{fienup1978reconstruction, fienup1982phase, Liu2015Vulnerability, shechtman2015phase}. Here, only amplitude information of the target is interested, which can be used as a constraint to significantly improve the speed and quality of reconstruction \cite{Ying2008two}.

To quantify the imaging system, the relationship between the field of view (FOV), the resolution and the spatial random phase modulator is analyzed.

The FOV of lensless Wiener-Khinchin telescope is limited by the memory effect range of the imaging system. Considering the height standard deviation $\omega$ of the SRPM, the normalized second-order correlation function of light fields between different incident angles without transverse translation is (see \href{link}{Supplement 1} for details)
\begin{equation}
\begin{split}
g_{\theta}^{(2)}\left({\Delta \theta}\right)
\approx \exp \left\{ { - {{\left( {\dfrac{{\pi n\omega }}{\lambda }\sin^2\left(\Delta \theta\right)} \right)}^2}} \right\},
\end{split}
\label{eq19}
\end{equation}
where $\Delta \theta$ is the variation of the incident angle. According to Eq. (\ref{eq19}), the FOV is proportional to $\dfrac{\lambda}{\omega}$.

In addition, Eq. (\ref{eq18}) leads to a limitation of the FOV,
\begin{equation}
FOV < \dfrac{L}{z_2}
\label{eq20}
\end{equation}
with $L$ denoting the CCD detector size. This equation indicates that FOV is also limited by the CCD detector size. In order to obtain a large FOV, the CCD detector size of lensless Wiener-Khinchin telescope is required to be much larger than $\dfrac{\lambda z_2}{\omega}$ in Eq. (\ref{eq19}).

Eq. (\ref{eq16}) indicates that the resolution not only depends on the aperture of the SRPM, but also the statistical properties of it. According to the convolution operation in $g_{h}^{(2)}\left(\dfrac{\Delta r_0}{2\lambda z_1}\right)$, we discuss two simple cases below, where the resolution is mainly limited by the aperture and the statistical properties of the SRPM, respectively.\\
\\
\textbf{Case 1}: resolution is mainly limited by the aperture.

When the full width at half maximum (FWHM) of $\small \exp \left\{-2\left[\dfrac{2\pi (n-1)}{\lambda}\right]^2\left[\omega^2-R_\eta\left(\dfrac{2\lambda z_1 z_2}{z_1+z_2}\nu\right)\right]\right\}$ is much smaller than the FWHM of $\mathcal{F}\left\{{\left| P\left(\mu+\mu_0\right)\right|^2}\right\}_{\mu \to \nu}$, we have
\begin{equation}
\begin{split}
g_{h}^{(2)}\left(\dfrac{\Delta r_0}{2\lambda z_1}\right)\approx\left|\mathcal{F}\left\{{\left| P\left(\mu\right)\right|^2}\right\}_{\mu \to -\dfrac{\Delta r_0}{2\lambda z_1}}\right|^2.
\end{split}
\label{eq21}
\end{equation}
For a circle aperture of the SRPM, $P(\mu)= \mathrm{circ} \left(\dfrac{\mu}{D}\right)$, and this leads to
\begin{equation}
\begin{split}
g_{h}^{(2)}\left(\dfrac{\Delta r_0}{2\lambda z_1}\right)\propto \left[{\dfrac{J_1(\dfrac{2\pi D\Delta r_0}{z_1\lambda})}{\dfrac{2\pi D\Delta r_0}{z_1\lambda}}}\right]^2.
\end{split}
\label{eq22}
\end{equation}

In this case, the resolution of the lensless Wiener-Khinchin telescope is proportional to $\lambda z_1/D$.\\
\\
\textbf{Case 2}: resolution is mainly limited by the statistical properties.

When the FWHM of $\exp \left\{-2\left[\dfrac{2\pi (n-1)}{\lambda}\right]^2\left[\omega^2-R_\eta\left(\dfrac{2\lambda z_1 z_2}{z_1+z_2}\nu\right)\right]\right\}$ is much larger than the FWHM of $\mathcal{F}\left\{{\left| P\left(\mu+\mu_0\right)\right|^2}\right\}_{\mu \to \nu}\quad$,
\begin{small}
\begin{equation}
\begin{split}
&g_{h}^{(2)}\left(\dfrac{\Delta r_0}{2\lambda z_1}\right)
\approx \exp \left\{-4\left[\dfrac{2\pi (n-1)\omega z_2\Delta r_0}{\lambda\left(z_1+z_2\right)\zeta}\right]^2\right\}\quad,
\end{split}
\label{eq23}
\end{equation}
\end{small}
with the first-order approximation. In this case, the resolution is proportional to $\left(1+\dfrac{z_1}{z_2}\right)\dfrac{\lambda\zeta}{(n-1)\omega}$.

For digital images, the reconstruction is also affected by the pixel size of the CCD detector. Due to Eq. (\ref{eq17}), the pixel size $P_{CCD}$ of the CCD detector is required by
\begin{equation}
P_{CCD}<\dfrac{z_2}{M z_1}g_{h}^{(2)}\left(\dfrac{\Delta r_0}{2\lambda z_1}\right),
\label{eq24}
\end{equation}
where $M$ denotes a split number for discrimination of resolution. 

\begin{figure}[htbp]
\centering
\includegraphics[width=0.8\linewidth]{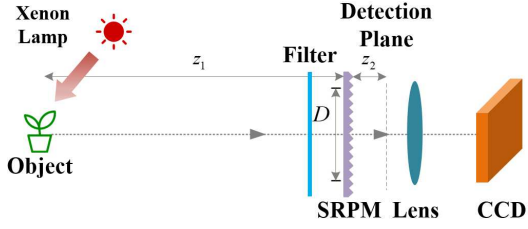}
\caption{Experimental setup of lensless Wiener-Khinchin telescope.}
\label{fig:ExperimentalSetup}
\end{figure}

The experimental setup is shown in Fig. \ref{fig:ExperimentalSetup}. An object is illuminated by a xenon lamp. The reflected light is filtered by a narrow-band filter, and modulated by a SRPM with a height standard deviation $\omega=1$ $\mu$m, a transverse correlation length $\zeta=44$ $\mu$m and the refractive index $n=1.46$, and then relayed by a lens with a magnification factor $\beta=10$ and a numerical aperture $N.A. = 0.25$ onto a CCD detector (APGCCD) with a pixel size 13 $\mu$m $\times$ 13 $\mu$m, which records the magnified intensity distribution. The lens is only used to amplify the intensity distribution to match the pixel size of the CCD detector, and is not necessary in some conditions.

\begin{figure}[htbp]
\centering
\includegraphics[width=5cm]{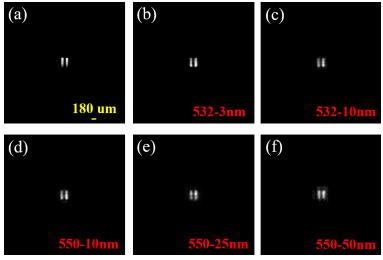}
\caption{Experimental results with different narrow-band filters. (a) A photograph of the double slit, where a yellow scale bar is inserted in the low right corner. Reconstructed images with different narrow-band filters: (b) $\lambda =532$ nm, $w=$ 3 nm, (c) $\lambda =532$ nm, $w=$ 10 nm, (d) $\lambda =550$ nm, $w=$ 10 nm, (e)$\lambda =550$ nm, $w=$ 25 nm, (f) $\lambda =550$ nm, $w=$ 50 nm.}
\label{fig:DoubleSlitExpLam}
\end{figure}

\begin{figure}[ht]
\centering
\includegraphics[width=8.5cm]{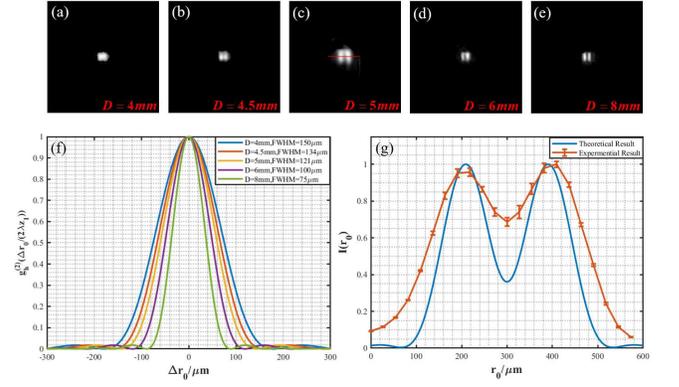}
\caption{Resolution at different apertures of the spatial random phase modulator. Reconstructed images with different apertures: (a) D= 4 mm, (b) D= 4.5 mm, (c) D= 5 mm, (d) D= 6 mm, (e) D= 8 mm. (f) The theoretical resolutions. (g) A comparison between theoretical and experimental resolutions at $D$= 5 mm, and the red line is a cross-section denoted by the dash line in Fig. \ref{fig:ExperimentalResultD}(c).}
\label{fig:ExperimentalResultD}
\end{figure}

\begin{figure}[ht]
\centering
\includegraphics[width=9cm]{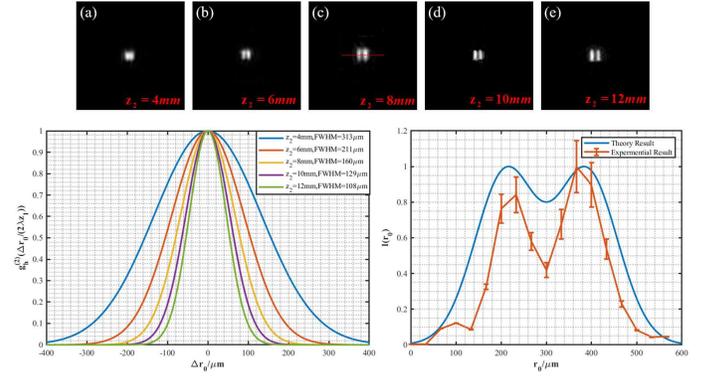}
\caption{Resolution at different $z_2$. Reconstructed images with different $z_2$: (a) $z_2=$ 4 mm, (b) $z_2=$ 6 mm, (c) $z_2=$ 8 mm, (d) $z_2=$ 10 mm, (e) $z_2=$ 12 mm. (f) The theoretical resolutions. (g) A comparison between theoretical and experimental resolutions at $z_2$= 8 mm, and the red line is a cross-section denoted by the dash line in Fig. \ref{fig:ExperimentalResultZ2}(c).}
\label{fig:ExperimentalResultZ2}
\end{figure}

In order to analyze the resolution of the system, a double slit (shown in Fig. \ref{fig:DoubleSlitExpLam}(a)) is selected. Since the image is obtained from the second-order spatial autocorrelation of thermal light, the temporal coherence is not strictly required. But the temporal coherence of the light field still affects the contrast of the spatial fluctuating pseudo-thermal light due to the dispersion of the spatial random phase modulator. The reflected light from the object is filtered by a narrow-band filter, whose central wavelength $\lambda$ is either 532 nm or 550 nm, and its bandwidth $w$ varies among 3 nm, 10 nm, 25 nm and 50 nm, when $z_1=$ 0.15 m, $z_2=$ 12 mm, and $D=$  8mm. The results with the same phase retrieval algorithm \cite{Liu2015Vulnerability} are shown in Fig. \ref{fig:DoubleSlitExpLam}. The experimental results show that the situation is better for narrow band light. In subsequent experiments, a narrow-band filter with a center wavelength $\lambda=$ 532 nm and a bandwidth $w=$ 10 nm is selected.

To verify Eq. (\ref{eq22}) in experiment, the aperture size is changed. Images with five different apertures $D=$ 4 mm, 4.5 mm, 5 mm, 6 mm, 8 mm are obtained, respectively, while $z_1=1.1$ m and $z_2=60$ mm are selected in accordance with \textbf{Case 1} (see Fig. \ref{fig:ExperimentalResultD}(a)-(e)). According to Eq. (\ref{eq21}), the theoretical resolutions with different apertures are shown in Fig. \ref{fig:ExperimentalResultD}(f), where FWHMs for $D=$ 4 mm, 4.5 mm, 5 mm, 6 mm and 8 mm are 150 $\mu$m, 134 $\mu$m, 121 $\mu$m, 100 $\mu$m and 75 $\mu$m, respectively. Fig. \ref{fig:ExperimentalResultD}(g) shows a comparison of theoretical and experimental resolutions at $D=$ 5 mm, where the red line is a cross-section denoted by the dash line in Fig. \ref{fig:ExperimentalResultD}(c). The experimental results show that the double slit can be distinguished at $D=5$ mm, which agrees well with theoretical results.

In \textbf{Case 2}, the resolution is mainly affected by the statistical properties of SRPM, which leads to a limitation of $z_2$ based on Eq. (\ref{eq23}). Five different $z_2$ (4 mm, 6 mm, 8 mm, 10 mm, 12 mm) are selected, and the reconstructed images are shown in Fig. \ref{fig:ExperimentalResultZ2}(a)-(e), while $z_1=0.3$ m and $D=8$ mm. The corresponding theoretical resolutions are shown in Fig. \ref{fig:ExperimentalResultZ2}(f), where FWHMs for $z_2=$ 4 mm, 6 mm, 8 mm, 10 mm and 12 mm are 313 $\mu$m, 211 $\mu$m, 160 $\mu$m, 129 $\mu$m and 108 $\mu$m, respectively. Fig. \ref{fig:ExperimentalResultZ2}(g) shows a comparison of theoretical and experimental resolutions at $z_2=$ 8 mm, where the red line is the cross-section denoted by the dash line in Fig. \ref{fig:ExperimentalResultZ2}(c). The results show that the double slit can be distinguished at $z_2=$ 8 mm.

\begin{figure}[htbp]
\centering
\includegraphics[height=1.6cm]{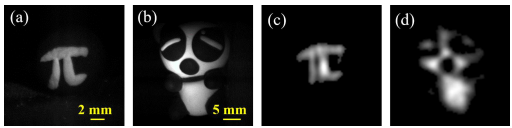}
\caption{Imaging a letter $\pi$ and a panda toy. (a) and (b) are photographs, where a yellow scale bar is inserted in the low right corner, respectively. (c) and (d) are reconstructed images, respectively.}
\label{fig:PiPandaExp}
\end{figure}

To further verify the imaging capability of lensless Wiener-Khinchin telescope, two targets, a letter $\pi$ and a panda toy, are imaged, respectively. Different system parameters are selected at $D=$ 8 mm. For the `$\pi$', $z_1=$ 0.5 m, $z_2=$ 2 mm, and for the `panda', $z_1=$ 1.5 m, $z_2=$ 3 mm. Both reconstructed images are shown in Fig. \ref{fig:PiPandaExp}.

For astronomical observations, the distance $z_1$ is nearly infinitely far away, which means $z_1\gg z_2$, so the resolution $g_{h}^{(2)}\left(\dfrac{\Delta r_0}{2\lambda z_1}\right)$ in Eq. (\ref{eq16}) is approximated to
\begin{small}
\begin{equation}
\begin{split}
&g_{h}^{(2)}\left(\dfrac{\Delta r_0}{2\lambda z_1}\right)\propto\left|\left\{{\exp \left\{-2\left[\dfrac{2\pi (n-1)}{\lambda}\right]^2\left[\omega^2-R_\eta\left(2\lambda z_2\nu\right)\right]\right\}}\right.\right.\\
&\qquad \qquad \qquad \left.\left.\otimes\mathcal{F}\left\{{\left| P\left(\mu\right)\right|^2}\right\}_{\mu \to \nu}\right\}_{\dfrac{\Delta r_0}{2\lambda z_1}}\right|^2.
\label{eq25}
\end{split}
\end{equation}
\end{small}

\begin{figure}[htbp]
\centering
\includegraphics[height=1.6cm]{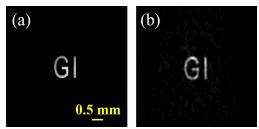}
\caption{Imaging an object placed equivalent infinite far away. (a) A photograph of the target, where a yellow scale bar is inserted in the low right corner, (b) reconstructed image.}
\label{fig:GIExp}
\end{figure}

A capital `GI' is placed on the focal plane of an optical lens before the SRPM to experimentally simulate the target placed infinite far away. The reconstructed image is shown in Fig. \ref{fig:GIExp}. Experimental results of Figs. \ref{fig:PiPandaExp} and \ref{fig:GIExp} prove the feasibility of lensless Wiener-Khinchin telescope in astronomical observations.

In conclusion, we present a theoretical framework for imaging schemes through random media, and propose lensless Wiener-Khinchin telescope based on high-order correlation of thermal light. The attempt to extract spatial information of an object from high-order correlation of light fields can be traced back to the famous HBT experiment in 1956 \cite{brown1956correlation, Smith2018Turbulence}, which is based on the second-order autocorrelation of light fields, and GI proposed in 1995, which is based on second-order mutual-correlation of light fields between the reference and test arms \cite{Strekalov1995Observation}. HBT experiment and many of the early works of GI \cite{Valencia2005Two, zhang2005correlated} perform ensemble statistics of the temporal fluctuating light field in time domain, which requires that the temporal resolution of the detector is close to or less than the coherence time of the light field \cite{liu2014lensless}. In contrast, by modulating true thermal light such as sunlight into a spatially fluctuating pseudo-thermal light field through a spatial random phase modulator \cite{liu2016spectral}, lensless Wiener-Khinchin telescope calculates the ensemble statistics of the spatially fluctuating pseudo-thermal light field in spatial domain, therefore, the detection of the temporal intensity fluctuation is not required. 

On the other hand, from the viewpoint of the intensity autocorrelation, single-shot imaging through scattering layers and around corners via speckle correlations presented by Katz \emph{et al.} \cite{katz2014non} didn't consider the effects of diffraction through the random phase modulation, and then the resolution of the imaging system can not be quantitatively calculated. By analyzing the spatial high-order correlation of light fields, the resolution is derived and experimentally verified in lensless Wiener-Khinchin telescope. The quantitative description of the imaging quality makes such imaging systems not only be demonstrated, but also be designed in practical applications.

Compared with lensless compressive sensing imaging \cite{Huang2013Lensless, Antipa2017DiffuserCam, Asif2017FlatCam} and lensless GI \cite{Chen2009Lensless, liu2014lensless, yu2016fourier}, neither a measurement matrix nor a calibration process is required. Thus lensless Wiener-Khinchin telescope has conspicuous advantages in applications such as X-ray astronomical observations, where the measurement matrix or the calibration for an unknown imaging distance is difficult and inaccurate. The cancellation of calibration also results in lower requirements in system stability. Moreover, considering the scattering media or the randomly inhomogeneous media as a spatial random phase modulator, lensless Wiener-Khinchin telescope may also open a door to quantitatively describe imaging through scattering media or randomly inhomogeneous media \cite{Choi2011Overcoming, Liutkus2014Imaging, Newman2014Imaging, Yilmaz2015Speckleb, liu2016spectral, Sahoo2017Single, Antipa2017DiffuserCam, Wang2015Ultra, Bertolotti2012Non, Katz2012Looking, katz2014non, Yang2014Imaging, labeyrie1970attainment, Gezari1972Speckle}.

\section*{Funding}
National Key Research and Development Program of China (2017YFB0503303). Hi-Tech Research and Development Program of China (2013AA122902 and 2013AA122901).

\begin{acknowledgments}
We thank Guowei Li and Guohai Situ for helpful discussions.
\end{acknowledgments}

\section*{References}


\begin{thebibliography}{46}%
\makeatletter
\providecommand \@ifxundefined [1]{%
 \@ifx{#1\undefined}
}%
\providecommand \@ifnum [1]{%
 \ifnum #1\expandafter \@firstoftwo
 \else \expandafter \@secondoftwo
 \fi
}%
\providecommand \@ifx [1]{%
 \ifx #1\expandafter \@firstoftwo
 \else \expandafter \@secondoftwo
 \fi
}%
\providecommand \natexlab [1]{#1}%
\providecommand \enquote  [1]{``#1''}%
\providecommand \bibnamefont  [1]{#1}%
\providecommand \bibfnamefont [1]{#1}%
\providecommand \citenamefont [1]{#1}%
\providecommand \href@noop [0]{\@secondoftwo}%
\providecommand \href [0]{\begingroup \@sanitize@url \@href}%
\providecommand \@href[1]{\@@startlink{#1}\@@href}%
\providecommand \@@href[1]{\endgroup#1\@@endlink}%
\providecommand \@sanitize@url [0]{\catcode `\\12\catcode `\$12\catcode
  `\&12\catcode `\#12\catcode `\^12\catcode `\_12\catcode `\%12\relax}%
\providecommand \@@startlink[1]{}%
\providecommand \@@endlink[0]{}%
\providecommand \url  [0]{\begingroup\@sanitize@url \@url }%
\providecommand \@url [1]{\endgroup\@href {#1}{\urlprefix }}%
\providecommand \urlprefix  [0]{URL }%
\providecommand \Eprint [0]{\href }%
\providecommand \doibase [0]{http://dx.doi.org/}%
\providecommand \selectlanguage [0]{\@gobble}%
\providecommand \bibinfo  [0]{\@secondoftwo}%
\providecommand \bibfield  [0]{\@secondoftwo}%
\providecommand \translation [1]{[#1]}%
\providecommand \BibitemOpen [0]{}%
\providecommand \bibitemStop [0]{}%
\providecommand \bibitemNoStop [0]{.\EOS\space}%
\providecommand \EOS [0]{\spacefactor3000\relax}%
\providecommand \BibitemShut  [1]{\csname bibitem#1\endcsname}%
\let\auto@bib@innerbib\@empty
\bibitem [{\citenamefont {Sheppard}(2017)}]{sheppard2017resolution}%
  \BibitemOpen
  \bibfield  {author} {\bibinfo {author} {\bibfnamefont {C.~J.~R.}\
  \bibnamefont {Sheppard}},\ }\href {\doibase 10.1002/jemt.22834} {\bibfield
  {journal} {\bibinfo  {journal} {Microscopy Research and Technique}\ }\textbf
  {\bibinfo {volume} {80}},\ \bibinfo {pages} {590} (\bibinfo {year}
  {2017})}\BibitemShut {NoStop}%
\bibitem [{\citenamefont {Baker}\ and\ \citenamefont
  {Kanade}(2002)}]{baker2002limits}%
  \BibitemOpen
  \bibfield  {author} {\bibinfo {author} {\bibfnamefont {S.}~\bibnamefont
  {Baker}}\ and\ \bibinfo {author} {\bibfnamefont {T.}~\bibnamefont {Kanade}},\
  }\href@noop {} {\bibfield  {journal} {\bibinfo  {journal} {IEEE Transactions
  on Pattern Analysis and Machine Intelligence}\ }\textbf {\bibinfo {volume}
  {24}},\ \bibinfo {pages} {1167} (\bibinfo {year} {2002})}\BibitemShut
  {NoStop}%
\bibitem [{\citenamefont {Roggemann}\ \emph {et~al.}(1997)\citenamefont
  {Roggemann}, \citenamefont {Welsh},\ and\ \citenamefont
  {Fugate}}]{Roggemann1997Improving}%
  \BibitemOpen
  \bibfield  {author} {\bibinfo {author} {\bibfnamefont {M.~C.}\ \bibnamefont
  {Roggemann}}, \bibinfo {author} {\bibfnamefont {B.~M.}\ \bibnamefont
  {Welsh}}, \ and\ \bibinfo {author} {\bibfnamefont {R.~Q.}\ \bibnamefont
  {Fugate}},\ }\href {\doibase 10.1103/revmodphys.69.437} {\bibfield  {journal}
  {\bibinfo  {journal} {Reviews of Modern Physics}\ }\textbf {\bibinfo {volume}
  {69}},\ \bibinfo {pages} {437} (\bibinfo {year} {1997})}\BibitemShut
  {NoStop}%
\bibitem [{\citenamefont {Hao}\ \emph {et~al.}(2013)\citenamefont {Hao},
  \citenamefont {Kuang}, \citenamefont {Gu}, \citenamefont {Wang},
  \citenamefont {Li}, \citenamefont {Ku}, \citenamefont {Li}, \citenamefont
  {Ge},\ and\ \citenamefont {Liu}}]{Hao2013microscopy}%
  \BibitemOpen
  \bibfield  {author} {\bibinfo {author} {\bibfnamefont {X.}~\bibnamefont
  {Hao}}, \bibinfo {author} {\bibfnamefont {C.}~\bibnamefont {Kuang}}, \bibinfo
  {author} {\bibfnamefont {Z.}~\bibnamefont {Gu}}, \bibinfo {author}
  {\bibfnamefont {Y.}~\bibnamefont {Wang}}, \bibinfo {author} {\bibfnamefont
  {S.}~\bibnamefont {Li}}, \bibinfo {author} {\bibfnamefont {Y.}~\bibnamefont
  {Ku}}, \bibinfo {author} {\bibfnamefont {Y.}~\bibnamefont {Li}}, \bibinfo
  {author} {\bibfnamefont {J.}~\bibnamefont {Ge}}, \ and\ \bibinfo {author}
  {\bibfnamefont {X.}~\bibnamefont {Liu}},\ }\href {\doibase
  10.1038/lsa.2013.64} {\bibfield  {journal} {\bibinfo  {journal} {Light:
  Science {\&} Applications}\ }\textbf {\bibinfo {volume} {2}},\ \bibinfo
  {pages} {e108} (\bibinfo {year} {2013})}\BibitemShut {NoStop}%
\bibitem [{\citenamefont {Choi}\ \emph {et~al.}(2011)\citenamefont {Choi},
  \citenamefont {Yang}, \citenamefont {Fang-Yen}, \citenamefont {Kang},
  \citenamefont {Lee}, \citenamefont {Dasari}, \citenamefont {Feld},\ and\
  \citenamefont {Choi}}]{Choi2011Overcoming}%
  \BibitemOpen
  \bibfield  {author} {\bibinfo {author} {\bibfnamefont {Y.}~\bibnamefont
  {Choi}}, \bibinfo {author} {\bibfnamefont {T.~D.}\ \bibnamefont {Yang}},
  \bibinfo {author} {\bibfnamefont {C.}~\bibnamefont {Fang-Yen}}, \bibinfo
  {author} {\bibfnamefont {P.}~\bibnamefont {Kang}}, \bibinfo {author}
  {\bibfnamefont {K.~J.}\ \bibnamefont {Lee}}, \bibinfo {author} {\bibfnamefont
  {R.~R.}\ \bibnamefont {Dasari}}, \bibinfo {author} {\bibfnamefont {M.~S.}\
  \bibnamefont {Feld}}, \ and\ \bibinfo {author} {\bibfnamefont
  {W.}~\bibnamefont {Choi}},\ }\href {\doibase 10.1103/physrevlett.107.023902}
  {\bibfield  {journal} {\bibinfo  {journal} {Physical Review Letters}\
  }\textbf {\bibinfo {volume} {107}} (\bibinfo {year} {2011}),\
  10.1103/physrevlett.107.023902}\BibitemShut {NoStop}%
\bibitem [{\citenamefont {Liutkus}\ \emph {et~al.}(2014)\citenamefont
  {Liutkus}, \citenamefont {Martina}, \citenamefont {Popoff}, \citenamefont
  {Chardon}, \citenamefont {Katz}, \citenamefont {Lerosey}, \citenamefont
  {Gigan}, \citenamefont {Daudet},\ and\ \citenamefont
  {Carron}}]{Liutkus2014Imaging}%
  \BibitemOpen
  \bibfield  {author} {\bibinfo {author} {\bibfnamefont {A.}~\bibnamefont
  {Liutkus}}, \bibinfo {author} {\bibfnamefont {D.}~\bibnamefont {Martina}},
  \bibinfo {author} {\bibfnamefont {S.}~\bibnamefont {Popoff}}, \bibinfo
  {author} {\bibfnamefont {G.}~\bibnamefont {Chardon}}, \bibinfo {author}
  {\bibfnamefont {O.}~\bibnamefont {Katz}}, \bibinfo {author} {\bibfnamefont
  {G.}~\bibnamefont {Lerosey}}, \bibinfo {author} {\bibfnamefont
  {S.}~\bibnamefont {Gigan}}, \bibinfo {author} {\bibfnamefont
  {L.}~\bibnamefont {Daudet}}, \ and\ \bibinfo {author} {\bibfnamefont
  {I.}~\bibnamefont {Carron}},\ }\href {\doibase 10.1038/srep05552} {\bibfield
  {journal} {\bibinfo  {journal} {Scientific Reports}\ }\textbf {\bibinfo
  {volume} {4}} (\bibinfo {year} {2014}),\ 10.1038/srep05552}\BibitemShut
  {NoStop}%
\bibitem [{\citenamefont {Newman}\ and\ \citenamefont
  {Webb}(2014)}]{Newman2014Imaging}%
  \BibitemOpen
  \bibfield  {author} {\bibinfo {author} {\bibfnamefont {J.~A.}\ \bibnamefont
  {Newman}}\ and\ \bibinfo {author} {\bibfnamefont {K.~J.}\ \bibnamefont
  {Webb}},\ }\href {\doibase 10.1103/physrevlett.113.263903} {\bibfield
  {journal} {\bibinfo  {journal} {Physical Review Letters}\ }\textbf {\bibinfo
  {volume} {113}} (\bibinfo {year} {2014}),\
  10.1103/physrevlett.113.263903}\BibitemShut {NoStop}%
\bibitem [{\citenamefont {Yilmaz}\ \emph {et~al.}(2015)\citenamefont {Yilmaz},
  \citenamefont {van Putten}, \citenamefont {Bertolotti}, \citenamefont
  {Lagendijk}, \citenamefont {Vos},\ and\ \citenamefont
  {Mosk}}]{Yilmaz2015Speckleb}%
  \BibitemOpen
  \bibfield  {author} {\bibinfo {author} {\bibfnamefont {H.}~\bibnamefont
  {Yilmaz}}, \bibinfo {author} {\bibfnamefont {E.~G.}\ \bibnamefont {van
  Putten}}, \bibinfo {author} {\bibfnamefont {J.}~\bibnamefont {Bertolotti}},
  \bibinfo {author} {\bibfnamefont {A.}~\bibnamefont {Lagendijk}}, \bibinfo
  {author} {\bibfnamefont {W.~L.}\ \bibnamefont {Vos}}, \ and\ \bibinfo
  {author} {\bibfnamefont {A.~P.}\ \bibnamefont {Mosk}},\ }\href {\doibase
  10.1364/optica.2.000424} {\bibfield  {journal} {\bibinfo  {journal} {Optica}\
  }\textbf {\bibinfo {volume} {2}},\ \bibinfo {pages} {424} (\bibinfo {year}
  {2015})}\BibitemShut {NoStop}%
\bibitem [{\citenamefont {Liu}\ \emph {et~al.}(2016)\citenamefont {Liu},
  \citenamefont {Tan}, \citenamefont {Wu}, \citenamefont {Li}, \citenamefont
  {Shen},\ and\ \citenamefont {Han}}]{liu2016spectral}%
  \BibitemOpen
  \bibfield  {author} {\bibinfo {author} {\bibfnamefont {Z.}~\bibnamefont
  {Liu}}, \bibinfo {author} {\bibfnamefont {S.}~\bibnamefont {Tan}}, \bibinfo
  {author} {\bibfnamefont {J.}~\bibnamefont {Wu}}, \bibinfo {author}
  {\bibfnamefont {E.}~\bibnamefont {Li}}, \bibinfo {author} {\bibfnamefont
  {X.}~\bibnamefont {Shen}}, \ and\ \bibinfo {author} {\bibfnamefont
  {S.}~\bibnamefont {Han}},\ }\href {\doibase 10.1038/srep25718} {\bibfield
  {journal} {\bibinfo  {journal} {Scientific Reports}\ }\textbf {\bibinfo
  {volume} {6}} (\bibinfo {year} {2016}),\ 10.1038/srep25718}\BibitemShut
  {NoStop}%
\bibitem [{\citenamefont {Sahoo}\ \emph {et~al.}(2017)\citenamefont {Sahoo},
  \citenamefont {Tang},\ and\ \citenamefont {Dang}}]{Sahoo2017Single}%
  \BibitemOpen
  \bibfield  {author} {\bibinfo {author} {\bibfnamefont {S.~K.}\ \bibnamefont
  {Sahoo}}, \bibinfo {author} {\bibfnamefont {D.}~\bibnamefont {Tang}}, \ and\
  \bibinfo {author} {\bibfnamefont {C.}~\bibnamefont {Dang}},\ }\href {\doibase
  10.1364/optica.4.001209} {\bibfield  {journal} {\bibinfo  {journal} {Optica}\
  }\textbf {\bibinfo {volume} {4}},\ \bibinfo {pages} {1209} (\bibinfo {year}
  {2017})}\BibitemShut {NoStop}%
\bibitem [{\citenamefont {Antipa}\ \emph {et~al.}(2017)\citenamefont {Antipa},
  \citenamefont {Kuo}, \citenamefont {Heckel}, \citenamefont {Mildenhall},
  \citenamefont {Bostan}, \citenamefont {Ng},\ and\ \citenamefont
  {Waller}}]{Antipa2017DiffuserCam}%
  \BibitemOpen
  \bibfield  {author} {\bibinfo {author} {\bibfnamefont {N.}~\bibnamefont
  {Antipa}}, \bibinfo {author} {\bibfnamefont {G.}~\bibnamefont {Kuo}},
  \bibinfo {author} {\bibfnamefont {R.}~\bibnamefont {Heckel}}, \bibinfo
  {author} {\bibfnamefont {B.}~\bibnamefont {Mildenhall}}, \bibinfo {author}
  {\bibfnamefont {E.}~\bibnamefont {Bostan}}, \bibinfo {author} {\bibfnamefont
  {R.}~\bibnamefont {Ng}}, \ and\ \bibinfo {author} {\bibfnamefont
  {L.}~\bibnamefont {Waller}},\ }\href {\doibase 10.1364/optica.5.000001}
  {\bibfield  {journal} {\bibinfo  {journal} {Optica}\ }\textbf {\bibinfo
  {volume} {5}},\ \bibinfo {pages} {1} (\bibinfo {year} {2017})}\BibitemShut
  {NoStop}%
\bibitem [{\citenamefont {Wang}\ and\ \citenamefont
  {Menon}(2015)}]{Wang2015Ultra}%
  \BibitemOpen
  \bibfield  {author} {\bibinfo {author} {\bibfnamefont {P.}~\bibnamefont
  {Wang}}\ and\ \bibinfo {author} {\bibfnamefont {R.}~\bibnamefont {Menon}},\
  }\href {\doibase 10.1364/optica.2.000933} {\bibfield  {journal} {\bibinfo
  {journal} {Optica}\ }\textbf {\bibinfo {volume} {2}},\ \bibinfo {pages} {933}
  (\bibinfo {year} {2015})}\BibitemShut {NoStop}%
\bibitem [{\citenamefont {Bertolotti}\ \emph {et~al.}(2012)\citenamefont
  {Bertolotti}, \citenamefont {van Putten}, \citenamefont {Blum}, \citenamefont
  {Lagendijk}, \citenamefont {Vos},\ and\ \citenamefont
  {Mosk}}]{Bertolotti2012Non}%
  \BibitemOpen
  \bibfield  {author} {\bibinfo {author} {\bibfnamefont {J.}~\bibnamefont
  {Bertolotti}}, \bibinfo {author} {\bibfnamefont {E.~G.}\ \bibnamefont {van
  Putten}}, \bibinfo {author} {\bibfnamefont {C.}~\bibnamefont {Blum}},
  \bibinfo {author} {\bibfnamefont {A.}~\bibnamefont {Lagendijk}}, \bibinfo
  {author} {\bibfnamefont {W.~L.}\ \bibnamefont {Vos}}, \ and\ \bibinfo
  {author} {\bibfnamefont {A.~P.}\ \bibnamefont {Mosk}},\ }\href {\doibase
  10.1038/nature11578} {\bibfield  {journal} {\bibinfo  {journal} {Nature}\
  }\textbf {\bibinfo {volume} {491}},\ \bibinfo {pages} {232} (\bibinfo {year}
  {2012})}\BibitemShut {NoStop}%
\bibitem [{\citenamefont {Katz}\ \emph {et~al.}(2012)\citenamefont {Katz},
  \citenamefont {Small},\ and\ \citenamefont {Silberberg}}]{Katz2012Looking}%
  \BibitemOpen
  \bibfield  {author} {\bibinfo {author} {\bibfnamefont {O.}~\bibnamefont
  {Katz}}, \bibinfo {author} {\bibfnamefont {E.}~\bibnamefont {Small}}, \ and\
  \bibinfo {author} {\bibfnamefont {Y.}~\bibnamefont {Silberberg}},\ }\href
  {\doibase 10.1038/nphoton.2012.150} {\bibfield  {journal} {\bibinfo
  {journal} {Nature Photonics}\ }\textbf {\bibinfo {volume} {6}},\ \bibinfo
  {pages} {549} (\bibinfo {year} {2012})}\BibitemShut {NoStop}%
\bibitem [{\citenamefont {Katz}\ \emph {et~al.}(2014)\citenamefont {Katz},
  \citenamefont {Heidmann}, \citenamefont {Fink},\ and\ \citenamefont
  {Gigan}}]{katz2014non}%
  \BibitemOpen
  \bibfield  {author} {\bibinfo {author} {\bibfnamefont {O.}~\bibnamefont
  {Katz}}, \bibinfo {author} {\bibfnamefont {P.}~\bibnamefont {Heidmann}},
  \bibinfo {author} {\bibfnamefont {M.}~\bibnamefont {Fink}}, \ and\ \bibinfo
  {author} {\bibfnamefont {S.}~\bibnamefont {Gigan}},\ }\href {\doibase
  10.1038/nphoton.2014.189} {\bibfield  {journal} {\bibinfo  {journal} {Nature
  Photonics}\ }\textbf {\bibinfo {volume} {8}},\ \bibinfo {pages} {784}
  (\bibinfo {year} {2014})}\BibitemShut {NoStop}%
\bibitem [{\citenamefont {Yang}\ \emph {et~al.}(2014)\citenamefont {Yang},
  \citenamefont {Pu},\ and\ \citenamefont {Psaltis}}]{Yang2014Imaging}%
  \BibitemOpen
  \bibfield  {author} {\bibinfo {author} {\bibfnamefont {X.}~\bibnamefont
  {Yang}}, \bibinfo {author} {\bibfnamefont {Y.}~\bibnamefont {Pu}}, \ and\
  \bibinfo {author} {\bibfnamefont {D.}~\bibnamefont {Psaltis}},\ }\href
  {\doibase 10.1364/oe.22.003405} {\bibfield  {journal} {\bibinfo  {journal}
  {Optics Express}\ }\textbf {\bibinfo {volume} {22}},\ \bibinfo {pages} {3405}
  (\bibinfo {year} {2014})}\BibitemShut {NoStop}%
\bibitem [{\citenamefont {Labeyrie}(1970)}]{labeyrie1970attainment}%
  \BibitemOpen
  \bibfield  {author} {\bibinfo {author} {\bibfnamefont {A.}~\bibnamefont
  {Labeyrie}},\ }\href@noop {} {\bibfield  {journal} {\bibinfo  {journal}
  {Astron. Astrophys.}\ }\textbf {\bibinfo {volume} {6}},\ \bibinfo {pages}
  {85} (\bibinfo {year} {1970})}\BibitemShut {NoStop}%
\bibitem [{\citenamefont {Gezari}\ \emph {et~al.}(1972)\citenamefont {Gezari},
  \citenamefont {Labeyrie},\ and\ \citenamefont
  {Stachnik}}]{Gezari1972Speckle}%
  \BibitemOpen
  \bibfield  {author} {\bibinfo {author} {\bibfnamefont {D.~Y.}\ \bibnamefont
  {Gezari}}, \bibinfo {author} {\bibfnamefont {A.}~\bibnamefont {Labeyrie}}, \
  and\ \bibinfo {author} {\bibfnamefont {R.~V.}\ \bibnamefont {Stachnik}},\
  }\href {\doibase 10.1086/180906} {\bibfield  {journal} {\bibinfo  {journal}
  {The Astrophysical Journal}\ }\textbf {\bibinfo {volume} {173}},\ \bibinfo
  {pages} {L1} (\bibinfo {year} {1972})}\BibitemShut {NoStop}%
\bibitem [{\citenamefont {Liu}\ \emph {et~al.}(2007)\citenamefont {Liu},
  \citenamefont {Cheng},\ and\ \citenamefont {Han}}]{Liu2007Ghost}%
  \BibitemOpen
  \bibfield  {author} {\bibinfo {author} {\bibfnamefont {H.}~\bibnamefont
  {Liu}}, \bibinfo {author} {\bibfnamefont {J.}~\bibnamefont {Cheng}}, \ and\
  \bibinfo {author} {\bibfnamefont {S.}~\bibnamefont {Han}},\ }\href {\doibase
  10.1063/1.2812597} {\bibfield  {journal} {\bibinfo  {journal} {Journal of
  Applied Physics}\ }\textbf {\bibinfo {volume} {102}},\ \bibinfo {pages}
  {103102} (\bibinfo {year} {2007})}\BibitemShut {NoStop}%
\bibitem [{\citenamefont {Zhang}\ \emph {et~al.}(2009)\citenamefont {Zhang},
  \citenamefont {Gong}, \citenamefont {Shen}, \citenamefont {Huang},\ and\
  \citenamefont {Han}}]{Zhang2009Improving}%
  \BibitemOpen
  \bibfield  {author} {\bibinfo {author} {\bibfnamefont {P.}~\bibnamefont
  {Zhang}}, \bibinfo {author} {\bibfnamefont {W.}~\bibnamefont {Gong}},
  \bibinfo {author} {\bibfnamefont {X.}~\bibnamefont {Shen}}, \bibinfo {author}
  {\bibfnamefont {D.}~\bibnamefont {Huang}}, \ and\ \bibinfo {author}
  {\bibfnamefont {S.}~\bibnamefont {Han}},\ }\href {\doibase
  10.1364/ol.34.001222} {\bibfield  {journal} {\bibinfo  {journal} {Optics
  Letters}\ }\textbf {\bibinfo {volume} {34}},\ \bibinfo {pages} {1222}
  (\bibinfo {year} {2009})}\BibitemShut {NoStop}%
\bibitem [{\citenamefont
  {Goodman}(2005{\natexlab{a}})}]{goodman2005introduction}%
  \BibitemOpen
  \bibfield  {author} {\bibinfo {author} {\bibfnamefont {J.~W.}\ \bibnamefont
  {Goodman}},\ }\href@noop {} {\emph {\bibinfo {title} {Introduction to Fourier
  optics}}},\ \bibinfo {number} {132-134}\ (\bibinfo  {publisher} {Roberts and
  Company Publishers},\ \bibinfo {year} {2005})\BibitemShut {NoStop}%
\bibitem [{\citenamefont {Cheng}\ and\ \citenamefont
  {Han}(2004)}]{cheng2004incoherent}%
  \BibitemOpen
  \bibfield  {author} {\bibinfo {author} {\bibfnamefont {J.}~\bibnamefont
  {Cheng}}\ and\ \bibinfo {author} {\bibfnamefont {S.}~\bibnamefont {Han}},\
  }\href {\doibase 10.1103/physrevlett.92.093903} {\bibfield  {journal}
  {\bibinfo  {journal} {Physical Review Letters}\ }\textbf {\bibinfo {volume}
  {92}} (\bibinfo {year} {2004}),\ 10.1103/physrevlett.92.093903}\BibitemShut
  {NoStop}%
\bibitem [{\citenamefont {Shapiro}\ and\ \citenamefont
  {Boyd}(2012)}]{Shapiro2012physics}%
  \BibitemOpen
  \bibfield  {author} {\bibinfo {author} {\bibfnamefont {J.~H.}\ \bibnamefont
  {Shapiro}}\ and\ \bibinfo {author} {\bibfnamefont {R.~W.}\ \bibnamefont
  {Boyd}},\ }\href {\doibase 10.1007/s11128-011-0356-5} {\bibfield  {journal}
  {\bibinfo  {journal} {Quantum Information Processing}\ }\textbf {\bibinfo
  {volume} {11}},\ \bibinfo {pages} {949} (\bibinfo {year} {2012})}\BibitemShut
  {NoStop}%
\bibitem [{\citenamefont {Goodman}(2007)}]{goodman2007speckle9}%
  \BibitemOpen
  \bibfield  {author} {\bibinfo {author} {\bibfnamefont {J.~W.}\ \bibnamefont
  {Goodman}},\ }\href@noop {} {\emph {\bibinfo {title} {Speckle phenomena in
  optics: theory and applications}}},\ \bibinfo {number} {9-12}\ (\bibinfo
  {publisher} {Roberts and Company Publishers},\ \bibinfo {year}
  {2007})\BibitemShut {NoStop}%
\bibitem [{\citenamefont {Goodman}(2015)}]{goodman2015statistical44}%
  \BibitemOpen
  \bibfield  {author} {\bibinfo {author} {\bibfnamefont {J.~W.}\ \bibnamefont
  {Goodman}},\ }\href@noop {} {\emph {\bibinfo {title} {Statistical optics}}},\
  \bibinfo {number} {44}\ (\bibinfo  {publisher} {John Wiley \& Sons},\
  \bibinfo {address} {New York},\ \bibinfo {year} {2015})\BibitemShut {NoStop}%
\bibitem [{\citenamefont {Feng}\ \emph {et~al.}(1988)\citenamefont {Feng},
  \citenamefont {Kane}, \citenamefont {Lee},\ and\ \citenamefont
  {Stone}}]{Feng1988Correlations}%
  \BibitemOpen
  \bibfield  {author} {\bibinfo {author} {\bibfnamefont {S.}~\bibnamefont
  {Feng}}, \bibinfo {author} {\bibfnamefont {C.}~\bibnamefont {Kane}}, \bibinfo
  {author} {\bibfnamefont {P.~A.}\ \bibnamefont {Lee}}, \ and\ \bibinfo
  {author} {\bibfnamefont {A.~D.}\ \bibnamefont {Stone}},\ }\href {\doibase
  10.1103/physrevlett.61.834} {\bibfield  {journal} {\bibinfo  {journal}
  {Physical Review Letters}\ }\textbf {\bibinfo {volume} {61}},\ \bibinfo
  {pages} {834} (\bibinfo {year} {1988})}\BibitemShut {NoStop}%
\bibitem [{\citenamefont {Osnabrugge}\ \emph {et~al.}(2017)\citenamefont
  {Osnabrugge}, \citenamefont {Horstmeyer}, \citenamefont {Papadopoulos},
  \citenamefont {Judkewitz},\ and\ \citenamefont
  {Vellekoop}}]{Osnabrugge2017Generalized}%
  \BibitemOpen
  \bibfield  {author} {\bibinfo {author} {\bibfnamefont {G.}~\bibnamefont
  {Osnabrugge}}, \bibinfo {author} {\bibfnamefont {R.}~\bibnamefont
  {Horstmeyer}}, \bibinfo {author} {\bibfnamefont {I.~N.}\ \bibnamefont
  {Papadopoulos}}, \bibinfo {author} {\bibfnamefont {B.}~\bibnamefont
  {Judkewitz}}, \ and\ \bibinfo {author} {\bibfnamefont {I.~M.}\ \bibnamefont
  {Vellekoop}},\ }\href {\doibase 10.1364/optica.4.000886} {\bibfield
  {journal} {\bibinfo  {journal} {Optica}\ }\textbf {\bibinfo {volume} {4}},\
  \bibinfo {pages} {886} (\bibinfo {year} {2017})}\BibitemShut {NoStop}%
\bibitem [{\citenamefont
  {Goodman}(2005{\natexlab{b}})}]{goodman2005introduction66}%
  \BibitemOpen
  \bibfield  {author} {\bibinfo {author} {\bibfnamefont {J.~W.}\ \bibnamefont
  {Goodman}},\ }\href@noop {} {\emph {\bibinfo {title} {Introduction to Fourier
  optics}}},\ \bibinfo {number} {66-67}\ (\bibinfo  {publisher} {Roberts and
  Company Publishers},\ \bibinfo {year} {2005})\BibitemShut {NoStop}%
\bibitem [{\citenamefont {Sinha}\ \emph {et~al.}(1988)\citenamefont {Sinha},
  \citenamefont {Sirota}, \citenamefont {Garoff},\ and\ \citenamefont
  {Stanley}}]{Sinha1988X}%
  \BibitemOpen
  \bibfield  {author} {\bibinfo {author} {\bibfnamefont {S.~K.}\ \bibnamefont
  {Sinha}}, \bibinfo {author} {\bibfnamefont {E.~B.}\ \bibnamefont {Sirota}},
  \bibinfo {author} {\bibfnamefont {S.}~\bibnamefont {Garoff}}, \ and\ \bibinfo
  {author} {\bibfnamefont {H.~B.}\ \bibnamefont {Stanley}},\ }\href {\doibase
  10.1103/physrevb.38.2297} {\bibfield  {journal} {\bibinfo  {journal}
  {Physical Review B}\ }\textbf {\bibinfo {volume} {38}},\ \bibinfo {pages}
  {2297} (\bibinfo {year} {1988})}\BibitemShut {NoStop}%
\bibitem [{\citenamefont {Cohen}(1998)}]{Cohen1998generalization}%
  \BibitemOpen
  \bibfield  {author} {\bibinfo {author} {\bibfnamefont {L.}~\bibnamefont
  {Cohen}},\ }in\ \href {\doibase 10.1109/icassp.1998.681753} {\emph {\bibinfo
  {booktitle} {Proceedings of the 1998 {IEEE} International Conference on
  Acoustics, Speech and Signal Processing, {ICASSP} 98 (Cat. No.98CH36181)}}}\
  (\bibinfo  {publisher} {{IEEE}},\ \bibinfo {year} {1998})\BibitemShut
  {NoStop}%
\bibitem [{\citenamefont
  {Goodman}(2005{\natexlab{c}})}]{goodman2005introduction8}%
  \BibitemOpen
  \bibfield  {author} {\bibinfo {author} {\bibfnamefont {J.~W.}\ \bibnamefont
  {Goodman}},\ }\href@noop {} {\emph {\bibinfo {title} {Introduction to Fourier
  optics}}},\ \bibinfo {number} {8-9}\ (\bibinfo  {publisher} {Roberts and
  Company Publishers},\ \bibinfo {year} {2005})\BibitemShut {NoStop}%
\bibitem [{\citenamefont {Fienup}(1978)}]{fienup1978reconstruction}%
  \BibitemOpen
  \bibfield  {author} {\bibinfo {author} {\bibfnamefont {J.~R.}\ \bibnamefont
  {Fienup}},\ }\href {\doibase 10.1364/ol.3.000027} {\bibfield  {journal}
  {\bibinfo  {journal} {Optics Letters}\ }\textbf {\bibinfo {volume} {3}},\
  \bibinfo {pages} {27} (\bibinfo {year} {1978})}\BibitemShut {NoStop}%
\bibitem [{\citenamefont {Fienup}(1982)}]{fienup1982phase}%
  \BibitemOpen
  \bibfield  {author} {\bibinfo {author} {\bibfnamefont {J.~R.}\ \bibnamefont
  {Fienup}},\ }\href {\doibase 10.1364/ao.21.002758} {\bibfield  {journal}
  {\bibinfo  {journal} {Applied Optics}\ }\textbf {\bibinfo {volume} {21}},\
  \bibinfo {pages} {2758} (\bibinfo {year} {1982})}\BibitemShut {NoStop}%
\bibitem [{\citenamefont {Liu}\ \emph {et~al.}(2015)\citenamefont {Liu},
  \citenamefont {Wu}, \citenamefont {He}, \citenamefont {Liao}, \citenamefont
  {Zhang},\ and\ \citenamefont {Peng}}]{Liu2015Vulnerability}%
  \BibitemOpen
  \bibfield  {author} {\bibinfo {author} {\bibfnamefont {X.}~\bibnamefont
  {Liu}}, \bibinfo {author} {\bibfnamefont {J.}~\bibnamefont {Wu}}, \bibinfo
  {author} {\bibfnamefont {W.}~\bibnamefont {He}}, \bibinfo {author}
  {\bibfnamefont {M.}~\bibnamefont {Liao}}, \bibinfo {author} {\bibfnamefont
  {C.}~\bibnamefont {Zhang}}, \ and\ \bibinfo {author} {\bibfnamefont
  {X.}~\bibnamefont {Peng}},\ }\href {\doibase 10.1364/oe.23.018955} {\bibfield
   {journal} {\bibinfo  {journal} {Optics Express}\ }\textbf {\bibinfo {volume}
  {23}},\ \bibinfo {pages} {18955} (\bibinfo {year} {2015})}\BibitemShut
  {NoStop}%
\bibitem [{\citenamefont {Shechtman}\ \emph {et~al.}(2015)\citenamefont
  {Shechtman}, \citenamefont {Eldar}, \citenamefont {Cohen}, \citenamefont
  {Chapman}, \citenamefont {Miao},\ and\ \citenamefont
  {Segev}}]{shechtman2015phase}%
  \BibitemOpen
  \bibfield  {author} {\bibinfo {author} {\bibfnamefont {Y.}~\bibnamefont
  {Shechtman}}, \bibinfo {author} {\bibfnamefont {Y.~C.}\ \bibnamefont
  {Eldar}}, \bibinfo {author} {\bibfnamefont {O.}~\bibnamefont {Cohen}},
  \bibinfo {author} {\bibfnamefont {H.~N.}\ \bibnamefont {Chapman}}, \bibinfo
  {author} {\bibfnamefont {J.}~\bibnamefont {Miao}}, \ and\ \bibinfo {author}
  {\bibfnamefont {M.}~\bibnamefont {Segev}},\ }\href {\doibase
  10.1109/msp.2014.2352673} {\bibfield  {journal} {\bibinfo  {journal} {{IEEE}
  Signal Processing Magazine}\ }\textbf {\bibinfo {volume} {32}},\ \bibinfo
  {pages} {87} (\bibinfo {year} {2015})}\BibitemShut {NoStop}%
\bibitem [{\citenamefont {Ying}\ \emph {et~al.}(2008)\citenamefont {Ying},
  \citenamefont {Wei}, \citenamefont {Shen},\ and\ \citenamefont
  {Han}}]{Ying2008two}%
  \BibitemOpen
  \bibfield  {author} {\bibinfo {author} {\bibfnamefont {G.}~\bibnamefont
  {Ying}}, \bibinfo {author} {\bibfnamefont {Q.}~\bibnamefont {Wei}}, \bibinfo
  {author} {\bibfnamefont {X.}~\bibnamefont {Shen}}, \ and\ \bibinfo {author}
  {\bibfnamefont {S.}~\bibnamefont {Han}},\ }\href {\doibase
  10.1016/j.optcom.2008.07.026} {\bibfield  {journal} {\bibinfo  {journal}
  {Optics Communications}\ }\textbf {\bibinfo {volume} {281}},\ \bibinfo
  {pages} {5130} (\bibinfo {year} {2008})}\BibitemShut {NoStop}%
\bibitem [{\citenamefont {Brown}\ and\ \citenamefont
  {Twiss}(1956)}]{brown1956correlation}%
  \BibitemOpen
  \bibfield  {author} {\bibinfo {author} {\bibfnamefont {R.~H.}\ \bibnamefont
  {Brown}}\ and\ \bibinfo {author} {\bibfnamefont {R.~Q.}\ \bibnamefont
  {Twiss}},\ }\href {\doibase 10.1038/177027a0} {\bibfield  {journal} {\bibinfo
   {journal} {Nature}\ }\textbf {\bibinfo {volume} {177}},\ \bibinfo {pages}
  {27} (\bibinfo {year} {1956})}\BibitemShut {NoStop}%
\bibitem [{\citenamefont {Smith}\ and\ \citenamefont
  {Shih}(2018)}]{Smith2018Turbulence}%
  \BibitemOpen
  \bibfield  {author} {\bibinfo {author} {\bibfnamefont {T.~A.}\ \bibnamefont
  {Smith}}\ and\ \bibinfo {author} {\bibfnamefont {Y.}~\bibnamefont {Shih}},\
  }\href {\doibase 10.1103/physrevlett.120.063606} {\bibfield  {journal}
  {\bibinfo  {journal} {Physical Review Letters}\ }\textbf {\bibinfo {volume}
  {120}} (\bibinfo {year} {2018}),\ 10.1103/physrevlett.120.063606}\BibitemShut
  {NoStop}%
\bibitem [{\citenamefont {Strekalov}\ \emph {et~al.}(1995)\citenamefont
  {Strekalov}, \citenamefont {Sergienko}, \citenamefont {Klyshko},\ and\
  \citenamefont {Shih}}]{Strekalov1995Observation}%
  \BibitemOpen
  \bibfield  {author} {\bibinfo {author} {\bibfnamefont {D.~V.}\ \bibnamefont
  {Strekalov}}, \bibinfo {author} {\bibfnamefont {A.~V.}\ \bibnamefont
  {Sergienko}}, \bibinfo {author} {\bibfnamefont {D.~N.}\ \bibnamefont
  {Klyshko}}, \ and\ \bibinfo {author} {\bibfnamefont {Y.~H.}\ \bibnamefont
  {Shih}},\ }\href {\doibase 10.1103/physrevlett.74.3600} {\bibfield  {journal}
  {\bibinfo  {journal} {Physical Review Letters}\ }\textbf {\bibinfo {volume}
  {74}},\ \bibinfo {pages} {3600} (\bibinfo {year} {1995})}\BibitemShut
  {NoStop}%
\bibitem [{\citenamefont {Valencia}\ \emph {et~al.}(2005)\citenamefont
  {Valencia}, \citenamefont {Scarcelli}, \citenamefont {D'Angelo},\ and\
  \citenamefont {Shih}}]{Valencia2005Two}%
  \BibitemOpen
  \bibfield  {author} {\bibinfo {author} {\bibfnamefont {A.}~\bibnamefont
  {Valencia}}, \bibinfo {author} {\bibfnamefont {G.}~\bibnamefont {Scarcelli}},
  \bibinfo {author} {\bibfnamefont {M.}~\bibnamefont {D'Angelo}}, \ and\
  \bibinfo {author} {\bibfnamefont {Y.}~\bibnamefont {Shih}},\ }\href {\doibase
  10.1103/physrevlett.94.063601} {\bibfield  {journal} {\bibinfo  {journal}
  {Physical Review Letters}\ }\textbf {\bibinfo {volume} {94}} (\bibinfo {year}
  {2005}),\ 10.1103/physrevlett.94.063601}\BibitemShut {NoStop}%
\bibitem [{\citenamefont {Zhang}\ \emph {et~al.}(2005)\citenamefont {Zhang},
  \citenamefont {Zhai}, \citenamefont {Wu},\ and\ \citenamefont
  {Chen}}]{zhang2005correlated}%
  \BibitemOpen
  \bibfield  {author} {\bibinfo {author} {\bibfnamefont {D.}~\bibnamefont
  {Zhang}}, \bibinfo {author} {\bibfnamefont {Y.~H.}\ \bibnamefont {Zhai}},
  \bibinfo {author} {\bibfnamefont {L.~A.}\ \bibnamefont {Wu}}, \ and\ \bibinfo
  {author} {\bibfnamefont {X.~H.}\ \bibnamefont {Chen}},\ }\href {\doibase
  10.1364/ol.30.002354} {\bibfield  {journal} {\bibinfo  {journal} {Optics
  Letters}\ }\textbf {\bibinfo {volume} {30}},\ \bibinfo {pages} {2354}
  (\bibinfo {year} {2005})}\BibitemShut {NoStop}%
\bibitem [{\citenamefont {Liu}\ \emph {et~al.}(2014)\citenamefont {Liu},
  \citenamefont {Chen}, \citenamefont {Yao}, \citenamefont {Yu}, \citenamefont
  {Zhai},\ and\ \citenamefont {Wu}}]{liu2014lensless}%
  \BibitemOpen
  \bibfield  {author} {\bibinfo {author} {\bibfnamefont {X.~F.}\ \bibnamefont
  {Liu}}, \bibinfo {author} {\bibfnamefont {X.~H.}\ \bibnamefont {Chen}},
  \bibinfo {author} {\bibfnamefont {X.~R.}\ \bibnamefont {Yao}}, \bibinfo
  {author} {\bibfnamefont {W.~K.}\ \bibnamefont {Yu}}, \bibinfo {author}
  {\bibfnamefont {G.~J.}\ \bibnamefont {Zhai}}, \ and\ \bibinfo {author}
  {\bibfnamefont {L.~A.}\ \bibnamefont {Wu}},\ }\href {\doibase
  10.1364/ol.39.002314} {\bibfield  {journal} {\bibinfo  {journal} {Optics
  Letters}\ }\textbf {\bibinfo {volume} {39}},\ \bibinfo {pages} {2314}
  (\bibinfo {year} {2014})}\BibitemShut {NoStop}%
\bibitem [{\citenamefont {Huang}\ \emph {et~al.}(2013)\citenamefont {Huang},
  \citenamefont {Jiang}, \citenamefont {Matthews},\ and\ \citenamefont
  {Wilford}}]{Huang2013Lensless}%
  \BibitemOpen
  \bibfield  {author} {\bibinfo {author} {\bibfnamefont {G.}~\bibnamefont
  {Huang}}, \bibinfo {author} {\bibfnamefont {H.}~\bibnamefont {Jiang}},
  \bibinfo {author} {\bibfnamefont {K.}~\bibnamefont {Matthews}}, \ and\
  \bibinfo {author} {\bibfnamefont {P.}~\bibnamefont {Wilford}},\ }in\ \href
  {\doibase 10.1109/icip.2013.6738433} {\emph {\bibinfo {booktitle} {2013
  {IEEE} International Conference on Image Processing}}}\ (\bibinfo
  {publisher} {{IEEE}},\ \bibinfo {year} {2013})\BibitemShut {NoStop}%
\bibitem [{\citenamefont {Asif}\ \emph {et~al.}(2017)\citenamefont {Asif},
  \citenamefont {Ayremlou}, \citenamefont {Sankaranarayanan}, \citenamefont
  {Veeraraghavan},\ and\ \citenamefont {Baraniuk}}]{Asif2017FlatCam}%
  \BibitemOpen
  \bibfield  {author} {\bibinfo {author} {\bibfnamefont {M.~S.}\ \bibnamefont
  {Asif}}, \bibinfo {author} {\bibfnamefont {A.}~\bibnamefont {Ayremlou}},
  \bibinfo {author} {\bibfnamefont {A.}~\bibnamefont {Sankaranarayanan}},
  \bibinfo {author} {\bibfnamefont {A.}~\bibnamefont {Veeraraghavan}}, \ and\
  \bibinfo {author} {\bibfnamefont {R.~G.}\ \bibnamefont {Baraniuk}},\ }\href
  {\doibase 10.1109/tci.2016.2593662} {\bibfield  {journal} {\bibinfo
  {journal} {{IEEE} Transactions on Computational Imaging}\ }\textbf {\bibinfo
  {volume} {3}},\ \bibinfo {pages} {384} (\bibinfo {year} {2017})}\BibitemShut
  {NoStop}%
\bibitem [{\citenamefont {Chen}\ \emph {et~al.}(2009)\citenamefont {Chen},
  \citenamefont {Liu}, \citenamefont {Luo},\ and\ \citenamefont
  {Wu}}]{Chen2009Lensless}%
  \BibitemOpen
  \bibfield  {author} {\bibinfo {author} {\bibfnamefont {X.~H.}\ \bibnamefont
  {Chen}}, \bibinfo {author} {\bibfnamefont {Q.}~\bibnamefont {Liu}}, \bibinfo
  {author} {\bibfnamefont {K.~H.}\ \bibnamefont {Luo}}, \ and\ \bibinfo
  {author} {\bibfnamefont {L.~A.}\ \bibnamefont {Wu}},\ }\href {\doibase
  10.1364/ol.34.000695} {\bibfield  {journal} {\bibinfo  {journal} {Optics
  Letters}\ }\textbf {\bibinfo {volume} {34}},\ \bibinfo {pages} {695}
  (\bibinfo {year} {2009})}\BibitemShut {NoStop}%
\bibitem [{\citenamefont {Yu}\ \emph {et~al.}(2016)\citenamefont {Yu},
  \citenamefont {Lu}, \citenamefont {Han}, \citenamefont {Xie}, \citenamefont
  {Du}, \citenamefont {Xiao},\ and\ \citenamefont {Zhu}}]{yu2016fourier}%
  \BibitemOpen
  \bibfield  {author} {\bibinfo {author} {\bibfnamefont {H.}~\bibnamefont
  {Yu}}, \bibinfo {author} {\bibfnamefont {R.}~\bibnamefont {Lu}}, \bibinfo
  {author} {\bibfnamefont {S.}~\bibnamefont {Han}}, \bibinfo {author}
  {\bibfnamefont {H.}~\bibnamefont {Xie}}, \bibinfo {author} {\bibfnamefont
  {G.}~\bibnamefont {Du}}, \bibinfo {author} {\bibfnamefont {T.}~\bibnamefont
  {Xiao}}, \ and\ \bibinfo {author} {\bibfnamefont {D.}~\bibnamefont {Zhu}},\
  }\href {\doibase 10.1103/physrevlett.117.113901} {\bibfield  {journal}
  {\bibinfo  {journal} {Physical Review Letters}\ }\textbf {\bibinfo {volume}
  {117}} (\bibinfo {year} {2016}),\ 10.1103/physrevlett.117.113901}\BibitemShut
  {NoStop}%
\end{thebibliography}
\end{document}